# Spin-orbit couplings of quantum fields in Schwarzschild spacetime

Zhi-Yong Wang *

*School of Optoelectronic Information, University of Electronic Science and Technology of China, Chengdu 610054, CHINA*

In Schwarzschild spacetime, the gravitational spin-orbit couplings of the massless Dirac field and the photon field can be studied in a unified way. In contrary to the previous investigations presented mainly at the quantum-mechanical level, our work is presented at the level of quantum field theory without resorting to the Foldy-Wouthuysen transformation. If massless Dirac particles and photons have the same momentums, their energy-level splittings due to the gravitational spin-orbit couplings are the same. Massless Dirac particles and photons coming from the Hawking radiations are partially polarized as long as their original momentums are not parallel to the radial direction of a Schwarzschild black hole.

There have been many investigations on the spin dynamics in a gravitational field [1-16], most of them are presented at the levels of quantum mechanics or classical field theory, rather than of quantum field theory. For example, some previous works are presented by studying the time evolution of single-particle variables (such as a spin vector or a four-dimensional (4D) velocity, *et al*), or by means of the semiclassical approximation based on the methods of the Foldy-Wouthuysen transformation [11-16]. In this paper, the gravitational spin-orbit couplings of the massless Dirac field and the photon field are studied in a unified way, from which one can obtain some new insights. We work throughout in geometrized units, $\hbar = c = G = 1$. The metric signature is $(-,+,+,+)$, and the 4D Minkowski spacetime metric tensor is denoted by $\eta_{\mu\nu} = \text{diag}(-1,1,1,1)$ ($\mu,\nu = 0,1,2,3$).

Complex conjugation is denoted by $*$ and hermitian conjugation by $\dagger$.

Outside a Schwarzschild black hole of mass $M$, the standard form of the Schwarzschild metric is

$$ds^2 = -(1-r_s/r)dt^2 + (1-r_s/r)^{-1}dr^2 + r^2(d\theta^2 + \sin^2\theta d\phi^2), \qquad (1)$$

where $r_s = 2M$ is the Schwarzschild radius of the black hole. One can introduce a new radial coordinate [17]

$$\rho = (r - r_s/2 + \sqrt{r^2 - r_s r})/2, \text{ or } r = \rho(1 + r_s/4\rho)^2, \qquad (2)$$

which implies that $\rho \to r_s/4$ for $r \to r_s$, and the isotropic form of the Schwarzschild metric reads:

$$ds^2 = -(1-r_s/4\rho)^2(1+r_s/4\rho)^{-2}dt^2 + (1+r_s/4\rho)^4(d\rho_1^2 + d\rho_2^2 + d\rho_3^2), \qquad (3)$$

where the variables $\rho_1$, $\rho_2$ and $\rho_3$ are defined by $\rho_1 = \rho\sin\theta\cos\phi$, $\rho_2 = \rho\sin\theta\sin\phi$, $\rho_3 = \rho\cos\theta$. Let $\boldsymbol{\rho} = (\rho_1, \rho_2, \rho_3)$, one has $\rho = |\boldsymbol{\rho}|$. To apply the connection coefficients in an orthonormal basis, let us rewrite Eq. (3) as [17]

$$ds^2 = -a_0^2 dt^2 + a_1^2 d\rho_1^2 + a_2^2 d\rho_2^2 + a_3^2 d\rho_3^2 = -(\theta^0)^2 + (\theta^1)^2 + (\theta^2)^2 + (\theta^3)^2, \qquad (4)$$

where

$$a_0 = (1-r_s/4\rho)(1+r_s/4\rho)^{-1}, \quad a = a_1 = a_2 = a_3 = (1+r_s/4\rho)^2, \qquad (5)$$

and $\theta^0 = a_0 dt$, $\theta^l = a_l d\rho^l$ ( $l = 1,2,3$ ) with $g_{\mu\nu} = \eta_{\mu\nu} = \text{diag}(-1,1,1,1)$ form an orthonormal basis, the dual basis is

$$e_\mu = a_\mu^{-1}\partial_\mu, \quad \mu = 0,1,2,3, \qquad (6)$$

here we have $x^\mu = (t, \boldsymbol{\rho})$, $\partial_\mu = \partial/\partial x^\mu = (\partial_0, \nabla)$, $\nabla = (\partial_1, \partial_2, \partial_3)$, $\partial_l = \partial/\partial \rho^l$, $l = 1,2,3$.

In the dual vector space there are the following commutators:

$$[e_\kappa, e_\lambda] = C^\mu{}_{\kappa\lambda} e_\mu, \quad C_{\mu\kappa\lambda} = \eta_{\mu\nu} C^\mu{}_{\kappa\lambda}, \qquad (7)$$

In the orthonormal basis the connection coefficients can be expressed in terms of the $C_{\mu\kappa\lambda}$,

$$\Gamma_{\kappa\lambda\mu} = -(C_{\kappa\lambda\mu} + C_{\lambda\mu\kappa} - C_{\mu\kappa\lambda})/2. \tag{8}$$

One can prove that ($\mu, \lambda = 0,1,2,3$, $l,m,n = 1,2,3$)

$$C_{\mu\lambda\lambda} = 0, \quad C_{llm} = -C_{lml} = a_m^{-1}\partial_m \ln a_l, \quad l \neq m, \tag{9-1}$$

$$C_{l0l} = -C_{ll0} = -a_0^{-1}\partial_0 \ln a_l, \quad C_{00l} = -C_{0l0} = -a_l^{-1}\partial_l \ln a_0, \tag{9-2}$$

$$C_{m0l} = -C_{ml0} = 0, \quad C_{0lm} = -C_{0ml} = 0, \quad C_{nlm} = -C_{nml} = 0, \quad l \neq m \neq n. \tag{9-3}$$

Let us first consider the Dirac field with vanishing rest mass. In Minkowski spacetime the free massless Dirac field satisfies the Dirac equation of $i\gamma^\mu \partial_\mu \varphi(x) = 0$, where $\gamma^\mu$'s are the Dirac matrices in Minkowski spacetime ($\gamma^\mu\gamma^\nu + \gamma^\nu\gamma^\mu = -2\eta^{\mu\nu}$). In curved spacetime the Dirac equation becomes (our conventions are different from those in Ref. [17])

$$i\gamma^\mu(e_\mu - i\Gamma_{\kappa\lambda\mu}S^{\kappa\lambda}/2)\varphi(x) = 0, \tag{10}$$

where $S^{\mu\nu} = i[\gamma^\mu, \gamma^\nu]/4$ is the 4D spin tensor of the Dirac field. In terms of the Pauli-matrix vector $\boldsymbol{\sigma} = (\sigma_1, \sigma_2, \sigma_3)$ and the 2×2 unit matrix $I_{2\times2}$, where

$$\sigma_1 = \begin{pmatrix} 0 & 1 \\ 1 & 0 \end{pmatrix}, \quad \sigma_2 = \begin{pmatrix} 0 & -i \\ i & 0 \end{pmatrix}, \quad \sigma_3 = \begin{pmatrix} 1 & 0 \\ 0 & -1 \end{pmatrix}, \tag{11}$$

one can define the Dirac matrices and other matrices in Minkowski spacetime as follows:

$$\gamma^0 = \begin{pmatrix} I_{2\times2} & 0 \\ 0 & -I_{2\times2} \end{pmatrix}, \quad \boldsymbol{\gamma} = \begin{pmatrix} 0 & \boldsymbol{\sigma} \\ -\boldsymbol{\sigma} & 0 \end{pmatrix}, \quad \boldsymbol{\alpha} = \begin{pmatrix} 0 & \boldsymbol{\sigma} \\ \boldsymbol{\sigma} & 0 \end{pmatrix}, \quad \boldsymbol{\Sigma} = \begin{pmatrix} \boldsymbol{\sigma} & 0 \\ 0 & \boldsymbol{\sigma} \end{pmatrix}. \tag{12}$$

Using $\sigma_l\sigma_m = i\varepsilon_{lmn}\sigma_n + \delta_{lm}$, where $\varepsilon_{klm} = \varepsilon^{klm}$ denote the totally antisymmetric tensor with $\varepsilon_{123} = 1$, $k,l,m = 1,2,3$, one can prove that

$$S_{l0} = i\alpha_l/2, \quad S_{lm} = \varepsilon_{lmn}\Sigma_n/2, \quad \gamma^l\Sigma^m = -\gamma^m\Sigma^l = i\varepsilon^{lmn}\gamma^n, \tag{13}$$

Using Eq. (9), $S^{\mu\nu} = -S^{\nu\mu}$ and $C_{\mu\kappa\lambda} = -C_{\mu\lambda\kappa}$, one can prove that

$$\gamma^\mu \Gamma_{\kappa\lambda\mu} S^{\kappa\lambda}/2 = -\gamma^0 S^{10} C_{00l} - \gamma^l S^{l0} C_{l0l} - (\gamma^2 S^{12} C_{221} - \gamma^3 S^{31} C_{331})$$
$$- (\gamma^3 S^{23} C_{332} - \gamma^1 S^{12} C_{112}) - (\gamma^1 S^{31} C_{113} - \gamma^2 S^{23} C_{223}) \quad . \tag{14}$$

Using Eqs. (9) and (13), Eq. (14) becomes

$$\gamma^\mu \Gamma_{\kappa\lambda\mu} \Sigma^{\kappa\lambda}/2 = -i(\gamma^0 a_0^{-1} \partial_0 \ln\sqrt{a_1 a_2 a_3} + \gamma^1 a_1^{-1} \partial_1 \ln\sqrt{a_0 a_2 a_3}$$
$$+ \gamma^2 a_2^{-1} \partial_2 \ln\sqrt{a_0 a_1 a_3} + \gamma^3 a_3^{-1} \partial_3 \ln\sqrt{a_0 a_1 a_2}) \quad . \tag{15}$$

Using Eqs. (6) and (15), seeing that $a = a_1 = a_2 = a_3$ and $\partial_0 a_\mu = 0$, Eq. (10) becomes

$$i\gamma^0 a_0^{-1} a \partial_0 \varphi + i\gamma^1 (\partial_1 - \partial_1 \ln\sqrt{a_0 a^2})\varphi + i\gamma^2 (\partial_2 - \partial_2 \ln\sqrt{a_0 a^2})\varphi$$
$$+ i\gamma^3 (\partial_3 - \partial_3 \ln\sqrt{a_0 a^2})\varphi = 0 \quad . \tag{16}$$

Define $\eta = a_0^{-1} a$, $\kappa = a_0 a^2$ (and then $\partial_0 \eta = \partial_0 \kappa = 0$), one can rewrite Eq. (16) as

$$i\gamma^0 \eta \partial_0 \varphi + i\gamma^1 (\partial_1 - \partial_1 \ln\sqrt{\kappa})\varphi + i\gamma^2 (\partial_2 - \partial_2 \ln\sqrt{\kappa})\varphi + i\gamma^3 (\partial_3 - \partial_3 \ln\sqrt{\kappa})\varphi = 0. \tag{17}$$

Define $\partial_0' = \eta \partial_0$, $\partial_\mu' = (\partial_0', \nabla) = (\eta \partial_0, \nabla)$, and,

$$\Pi_D^\mu = \partial_\mu' \ln\sqrt{\kappa} = (\Pi_D^0, \mathbf{\Pi}_D), \quad \Pi_D^0 = 0, \quad \mathbf{\Pi}_D = \nabla \ln\sqrt{\kappa}. \tag{18}$$

Using Eq. (18) one can rewrite Eq. (17) as, formally

$$i\gamma^\mu (\partial_\mu' - \Pi_{D\mu})\varphi(x) = 0. \tag{19}$$

Eq. (19) is valid for the Dirac field only in Schwarzschild spacetime. We call $\Pi_D^\mu = \partial_\mu' \ln\sqrt{\kappa}$ the pseudo-gauge potential for the Dirac field in Schwarzschild spacetime. BTW, using Eq. (19) one can obtain the equation of continuity:

$$(\partial_\mu' - 2\Pi_{D\mu}) j^\mu = 0, \tag{20}$$

where $j^\mu = \bar{\varphi} \gamma^\mu \varphi$, $\bar{\varphi} = \varphi^\dagger \gamma^0$. Using $\partial_0' = \eta \partial/\partial t$, $\nabla \eta = 2\eta \Lambda$ with $\Lambda = \nabla \ln\sqrt{\eta}$, it is easy to show that

$$\nabla \partial_0' f - \partial_0' \nabla f = 2\Lambda \partial_0' f, \quad \nabla \times \Lambda f = -\Lambda \times \nabla f, \quad \partial_0' \Lambda f = \Lambda \partial_0' f. \tag{21}$$

Let $A_\mu = (A_0, \mathbf{A})$, $B_\mu = (B_0, \mathbf{B})$, using $(\boldsymbol{\sigma} \cdot \mathbf{A})(\boldsymbol{\sigma} \cdot \mathbf{B}) = \mathbf{A} \cdot \mathbf{B} + i\boldsymbol{\sigma} \cdot (\mathbf{A} \times \mathbf{B})$ and Eq. (12) one has

$$(\gamma^{\mu} A_{\mu})(\gamma^{\nu} B_{\nu}) = -A^{\mu} B_{\mu} - i\boldsymbol{\Sigma} \cdot (\boldsymbol{A} \times \boldsymbol{B}) + \boldsymbol{\alpha} \cdot (A_0 \boldsymbol{B} - \boldsymbol{A} B_0). \quad (22)$$

Using Eqs. (21) and (22), it follows from Eq. (19) that

$$(\partial'^{\mu} - \Pi_D^{\mu})(\partial'_{\mu} - \Pi_{D\mu})\varphi + 2\boldsymbol{\alpha} \cdot \boldsymbol{\Lambda} \partial'_0 \varphi = 0. \quad (23)$$

Let $\Phi = \kappa^{-1/2} \varphi$, it follows from Eq. (11) that

$$i\gamma^{\mu} \partial'_{\mu} \Phi(x) = 0. \quad (24)$$

The form of Eq. (24) is similar to the one of $i\gamma^{\mu} \partial_{\mu} \varphi(x) = 0$ in Minkowski spacetime, but for the velocity of light in Minkowski vacuum being replaced with the one in an equivalent medium with the refractive index of $\eta = a_0^{-1} a$. According to the tetrad formalism, the metric tensor $g_{\mu\nu}$ for Riemannian spacetime transforms like a scalar with respect to Lorentz transformations in tangent space (being equivalent to a Minkowski spacetime locally). As a result, one can show that $\eta = a_0^{-1} a$ and $\kappa = a_0 a^2$ are Lorentz scalars in the local Minkowski spacetime.

Now, let us consider the photon field in Schwarzschild spacetime. In our work, the photon field just plays the role of a matter field which interacts through a gravitational field, and then we ignore its role as a gauge field mediating the interaction between the matter fields with electric charges, and let the charge and current densities vanish. For the moment, the photon field can be described via a field quantity that corresponds to the (1, 0)+(0, 1) spinor representation of the Lorentz group. Such field quantity can be defined in terms of the column-matrix form of the electromagnetic field intensities $\boldsymbol{E} = (E_1, E_2, E_3)$ and $\boldsymbol{H} = (H_1, H_2, H_3)$, i.e.,

$$\psi(x) = \frac{1}{\sqrt{2}} \begin{pmatrix} \boldsymbol{E} \\ i\boldsymbol{H} \end{pmatrix}, \quad \boldsymbol{E} = \begin{pmatrix} E_1 \\ E_2 \\ E_3 \end{pmatrix}, \quad \boldsymbol{H} = \begin{pmatrix} H_1 \\ H_2 \\ H_3 \end{pmatrix}. \quad (25)$$

Maxwell's equations in Minkowski vacuum can be rewritten as the Dirac-like equation [18]

$$i\beta^{\mu}\partial_{\mu}\psi(x)=0, \qquad (26)$$

where, by means of the $3\times 3$ unit matrix $I_{3\times 3}$ and the matrix vector $\boldsymbol{\tau}=(\tau_1,\tau_2,\tau_3)$ with the matrix components

$$\tau_1=\begin{pmatrix}0&0&0\\0&0&-i\\0&i&0\end{pmatrix},\ \tau_2=\begin{pmatrix}0&0&i\\0&0&0\\-i&0&0\end{pmatrix},\ \tau_3=\begin{pmatrix}0&-i&0\\i&0&0\\0&0&0\end{pmatrix}, \qquad (27)$$

one can define the matrices $\beta^{\mu}=(\beta^0,\boldsymbol{\beta})$, $\boldsymbol{\alpha}=\beta^0\boldsymbol{\beta}=-\boldsymbol{\beta}\beta^0$ and $\Sigma$ as follows:

$$\beta^0=\begin{pmatrix}I_{3\times 3}&0\\0&-I_{3\times 3}\end{pmatrix},\ \boldsymbol{\beta}=\begin{pmatrix}0&\boldsymbol{\tau}\\-\boldsymbol{\tau}&0\end{pmatrix},\ \boldsymbol{\alpha}_p=\begin{pmatrix}0&\boldsymbol{\tau}\\\boldsymbol{\tau}&0\end{pmatrix},\ \Sigma_p=\begin{pmatrix}\boldsymbol{\tau}&0\\0&\boldsymbol{\tau}\end{pmatrix}. \qquad (28)$$

Under an infinitesimal Lorentz transformation $x^{\mu}\to x'^{\mu}=x^{\mu}-\varepsilon^{\mu\nu}x_{\nu}$ ($\varepsilon_{\mu\nu}$ is a real infinitesimal antisymmetric tensor), the photon field $\psi(x)$ transforms in the way [18],

$$\psi(x)\to\psi'(x')=(1-i\varepsilon_{\mu\nu}\Sigma^{\mu\nu}/2)\psi(x), \qquad (29)$$

where

$$\Sigma_{lm}=\varepsilon_{lmn}\Sigma_p^n,\ \Sigma^{0l}=i\alpha_p^l,\ l,m,n=1,2,3, \qquad (30)$$

where $\Sigma^{\mu\nu}$ is the infinitesimal generator of Lorentz group (corresponding to the (1, 0)+(0, 1) spinor representation), i.e., the 4D spin tensor of the photon field; $\varepsilon_{lmn}=\varepsilon^{lmn}$ denotes the full antisymmetric tensor with $\varepsilon_{123}=1$, $\Sigma_p^n$ and $\alpha_p^l$ are given by Eq. (28). For the moment, the Dirac-like equation (26) is Lorentz covariant. Moreover, in terms of the (1, 0)+(0, 1) spinor $\psi(x)$, one can discuss the quantization of the photon field easily [18].

Likewise, based on spin connection and the tetrad formalism, one can show that in curved spacetime the Dirac-like equation becomes

$$i\beta^{\mu}(e_{\mu}-i\Gamma_{\kappa\lambda\mu}\Sigma^{\kappa\lambda}/2)\psi(x)=0. \qquad (31)$$

Using Eq. (9), $\Sigma^{\mu\nu}=-\Sigma^{\nu\mu}$ and $C_{\mu\kappa\lambda}=-C_{\mu\lambda\kappa}$, one can prove that

$$\beta^{\mu}\Gamma_{\kappa\lambda\mu}\Sigma^{\kappa\lambda}/2 = -\beta^{0}\Sigma^{l0}C_{00l} - \beta^{l}\Sigma^{l0}C_{l0l} - (\beta^{2}\Sigma^{12}C_{221} - \beta^{3}\Sigma^{31}C_{331})$$
$$- (\beta^{3}\Sigma^{23}C_{332} - \beta^{1}\Sigma^{12}C_{112}) - (\beta^{1}\Sigma^{31}C_{113} - \beta^{2}\Sigma^{23}C_{223}) \quad . \tag{32}$$

Using $a = a_1 = a_2 = a_3$ and Eq. (9), Eq. (32) becomes

$$\beta^{\mu}\Gamma_{\kappa\lambda\mu}\Sigma^{\kappa\lambda}/2 = \beta_{l}\Sigma^{l0}a_{0}^{-1}\partial_{t}\ln a + \beta^{0}\Sigma^{m0}a^{-1}\partial_{m}\ln a_{0} - (\beta^{2}\Sigma^{12} - \beta^{3}\Sigma^{31})a^{-1}\partial_{1}\ln a$$
$$- (\beta^{3}\Sigma^{23} - \beta^{1}\Sigma^{12})a^{-1}\partial_{2}\ln a - (\beta^{1}\Sigma^{31} - \beta^{2}\Sigma^{23})a^{-1}\partial_{3}\ln a \quad . \tag{33}$$

Using Eqs. (27) and (28), $[\tau_l, \tau_m] = i\varepsilon_{lmn}\tau_n$, one can prove that

$$\beta^{l}\Sigma^{m}_{p} - \beta^{m}\Sigma^{l}_{p} = i\varepsilon^{lmn}\beta_{n}, \quad \beta^{l}\Sigma^{m}_{p} - \Sigma^{m}_{p}\beta^{l} = i\varepsilon^{lmn}\beta_{n}. \tag{34}$$

Using $\Sigma_{lm} = \varepsilon_{lmn}\Sigma^{n}_{p}$, $\Sigma^{0l} = i\alpha^{l}_{p}$, $\beta^{l} = \beta^{0}\alpha^{l}_{p}$, $\alpha_{p}\cdot\alpha_{p} = 2$ and Eq. (34), Eq. (33) becomes

$$\beta^{\mu}\Gamma_{\kappa\lambda\mu}\Sigma^{\kappa\lambda}/2 = -ia_{0}^{-1}\beta^{0}\partial_{t}\ln a^{2} - ia^{-1}\beta^{l}\partial_{l}\ln a_{0}a. \tag{35}$$

Substituting Eqs. (6) and (35) into Eq. (31), using $a = a_1 = a_2 = a_3$ one has

$$i\beta^{0}a_{0}^{-1}a(\partial_{t} - \partial_{t}\ln a^{2})\psi + i\beta^{1}(\partial_{1} - \partial_{1}\ln a_{0}a)\psi$$
$$+ i\beta^{2}(\partial_{2} - \partial_{2}\ln a_{0}a)\psi + i\beta^{3}(\partial_{3} - \partial_{3}\ln a_{0}a)\psi = 0 \quad . \tag{36}$$

Because of $\partial_{0}\ln a^{2} = \partial_{0}\ln a_{0}a = 0$, Eq. (36) can be rewritten as, formally

$$i\beta^{0}\partial'_{0}\psi(x) + i\beta^{l}(\partial_{l} - \Pi_{l})\psi(x) = 0, \text{ or } i\beta^{\mu}(\partial'_{\mu} - \Pi_{\mu})\psi(x) = 0, \tag{37}$$

where $\partial'_{\mu} = (\eta\partial_{t}, \nabla)$ with $\eta = a_{0}^{-1}a$, and

$$\Pi^{\mu} = \partial'_{\mu}\ln\sqrt{\varpi} = (\Pi^{0}, \boldsymbol{\Pi}), \quad \Pi^{0} = 0, \quad \boldsymbol{\Pi} = \nabla\ln\sqrt{\varpi}, \quad \varpi = (a_{0}a)^{2}. \tag{38}$$

We call $\Pi^{\mu} = \partial'_{\mu}\ln\sqrt{\varpi}$ the pseudo-gauge potential for the photon field in Schwarzschild spacetime. Likewise, according to the tetrad formalism, one can show that $\eta = a_{0}^{-1}a$ and $\varpi = (a_{0}a)^{2}$ are Lorentz scalars in tangent space (a local Minkowski spacetime). To guarantee Eq. (37) be Lorentz covariant in the local Minkowski spacetime, the transversality conditions should become

$$(\nabla - \boldsymbol{\Pi})\cdot\boldsymbol{E} = 0, \quad (\nabla - \boldsymbol{\Pi})\cdot\boldsymbol{H} = 0. \tag{39}$$

Similarly, Let $\Psi = \varpi^{-1/2}\psi = (\boldsymbol{E}' \quad i\boldsymbol{H}')^{T}/\sqrt{2}$ (and then $\boldsymbol{E}' = \varpi^{-1/2}\boldsymbol{E}$, $\boldsymbol{H}' = \varpi^{-1/2}\boldsymbol{H}$, the

superscript T denotes the matrix transpose), one can show that Eq. (37) can be rewritten as

$$i\beta^\mu \partial'_\mu \Psi = 0.  \tag{40}$$

For the moment Eq. (39) becomes

$$\nabla \cdot \bm{E}' = \nabla \cdot \bm{H}' = 0.  \tag{41}$$

Eqs. (40) and (41) imply that, with the velocity of light in Minkowski vacuum replaced with the one in the equivalent medium (with the refractive index $\eta = a_0^{-1} a$), the Dirac-like equation in Schwarzschild spacetime is equivalent to the one in Minkowski spacetime.

Let $a^\mu = (a^0, \bm{a})$ and $b^\mu = (b^0, \bm{b})$ be two 4D vectors, the column matrix forms of $\bm{a}$ and $\bm{b}$ are denoted as $\bm{a}_m = \begin{pmatrix} a_1 & a_2 & a_3 \end{pmatrix}^T$, $\bm{b}_m = \begin{pmatrix} b_1 & b_2 & b_3 \end{pmatrix}^T$ (the superscript T denotes the matrix transpose, the same below), respectively, one can prove that

$$(\bm{\tau} \cdot \bm{a})(\bm{\tau} \cdot \bm{b}) = \bm{a} \cdot \bm{b} I_{3\times 3} + i\bm{\tau} \cdot (\bm{a} \times \bm{b}) - \bm{a}_m \bm{b}_m^T,  \tag{42}$$

$$(\beta^\mu a_\mu)(\beta^\nu b_\nu) = -a^\mu b_\mu I_{6\times 6} - i\bm{\Sigma}_p \cdot (\bm{a} \times \bm{b}) + \bm{\alpha}_p \cdot (\bm{a} b^0 - a^0 \bm{b}) + I_{2\times 2} \otimes \bm{a}_m \bm{b}_m^T.  \tag{43}$$

Using Eqs. (21), (37)-(39), (42), (43) and $\bm{\Lambda} = \nabla \ln \sqrt{\eta}$, one can obtain

$$(\partial'^\mu - \Pi^\mu)(\partial'_\mu - \Pi_\mu)\psi + 2\bm{\alpha}_p \cdot \bm{\Lambda} \partial'_0 \psi = 0.  \tag{44}$$

It is difficult to solve the exact solutions of Eqs. (24) and Eq. (40), but we can study the dispersion relations of massless Dirac particles and photons in Schwarzschild spacetime. Substituting $\Phi = \begin{pmatrix} \chi & \zeta \end{pmatrix}^T$ and $\Psi = \begin{pmatrix} \bm{E}' & i\bm{H}' \end{pmatrix}^T / \sqrt{2}$ into Eqs. (24) and (40), one can obtain, respectively,

$$(\bm{\sigma} \cdot \nabla)\chi = -\eta \partial_t \zeta, \quad (\bm{\sigma} \cdot \nabla)\zeta = -\eta \partial_t \chi.  \tag{45}$$

$$(\bm{\tau} \cdot \nabla)\bm{H}' = i\eta \partial_t \bm{E}', \quad (\bm{\tau} \cdot \nabla)\bm{E}' = -i\eta \partial_t \bm{H}'.  \tag{46}$$

where $\bm{\sigma} = (\sigma_1, \sigma_2, \sigma_3)$ is the Pauli matrix vector given by Eq. (11), and the matrix vector $\bm{\tau} = (\tau_1, \tau_2, \tau_3)$ is given by Eq. (27). It follows from Eqs. (41), (45) and (46) that,

$$\eta^2 \partial_t^2 f = -\nabla^2 f + 2\Lambda \cdot \nabla f + 2\mathrm{i} \boldsymbol{s} \cdot (\Lambda \times \nabla f), \tag{47}$$

where $\Lambda = \nabla \ln \sqrt{\eta}$, and

$$\boldsymbol{s} = \begin{cases} \boldsymbol{\sigma}, & \text{for } f(t,\boldsymbol{\rho}) = \zeta, \chi \\ \boldsymbol{\tau}, & \text{for } f(t,\boldsymbol{\rho}) = \boldsymbol{E'}, \boldsymbol{H'} \end{cases}. \tag{48}$$

The last term on the right-hand side of Eq. (47) represents the spin-orbit coupling interaction. Let $\Lambda = |\Lambda|$, $\boldsymbol{e}_\rho \equiv \boldsymbol{\rho}/\rho$, using Eq. (5) and $\eta = a_0^{-1} a$, one has

$$\eta = (1 - r_s/4\rho)^{-1}(1 + r_s/4\rho)^3, \tag{49}$$

$$\boldsymbol{\Lambda} = -\frac{r_s \boldsymbol{e}_\rho}{8\rho^2}[\frac{3}{(1+r_s/4\rho)} + \frac{1}{(1-r_s/4\rho)}] = -\Lambda \boldsymbol{e}_\rho. \tag{50}$$

Seeing that there is a spherical symmetry and $u = 1/\eta$ is the velocity of light in the equivalent medium, for distant observers let us assume that

$$f(t,\boldsymbol{\rho}) = A_0(\rho)\exp[-\mathrm{i}(\omega t - \eta \boldsymbol{k} \cdot \boldsymbol{\rho})] = A(\rho,t)\exp(\mathrm{i}\eta \boldsymbol{k} \cdot \boldsymbol{\rho}), \tag{51}$$

where $A(\rho,t) = A_0(\rho)\exp(-\mathrm{i}\omega t)$, $\partial_t A_0(\rho) = 0$, $\partial_t \omega = 0$, $\partial_\mu k = 0$, $A_0 = A_0(\rho)$ and $\omega = \omega(\rho)$ are the functions of $\rho = |\boldsymbol{\rho}|$. Let $\boldsymbol{\Gamma} \equiv \nabla \ln A(\rho,t) = \boldsymbol{e}_\rho[\partial A(\rho,t)/\partial \rho]$, $\boldsymbol{k} \cdot \boldsymbol{e}_\rho = |\boldsymbol{k}|\cos\Theta$, $\boldsymbol{e}_\rho \times \boldsymbol{k} = |\boldsymbol{k}|\sin\Theta \boldsymbol{e}_t$, where $\Theta$ is the included angle between the unit vector of $\boldsymbol{e}_\rho$ and the vector of $\boldsymbol{k}$, $\boldsymbol{e}_t$ is a unit vector perpendicular to $\boldsymbol{e}_\rho$ and $\boldsymbol{k}$, using Eqs. (49)-(51), one can prove that Eq. (47) becomes

$$\begin{aligned}\eta^2 \omega^2 f &= \eta^2 k^2 f + 2\boldsymbol{\Lambda} \cdot \boldsymbol{\Gamma} f + \boldsymbol{\Gamma}^2 f + \nabla \cdot \boldsymbol{\Gamma} f + 2\eta|\boldsymbol{k}|\Lambda \sin\Theta(\boldsymbol{s} \cdot \boldsymbol{e}_t)f \\ &+ 4\eta^2(\boldsymbol{k} \cdot \boldsymbol{\rho})^2 \Lambda^2 f + 4\eta^2(\boldsymbol{k} \cdot \boldsymbol{\rho})(\boldsymbol{k} \cdot \boldsymbol{\Lambda})f - 2\mathrm{i}\eta(\boldsymbol{k} \cdot \boldsymbol{\rho})(\nabla \cdot \boldsymbol{\Lambda})f \\ &- 2\mathrm{i}\eta(\boldsymbol{k} \cdot \boldsymbol{\Lambda})f + 2\mathrm{i}\eta(\boldsymbol{k} \cdot \boldsymbol{\rho})(\boldsymbol{\Gamma} \cdot \boldsymbol{\Lambda})f + \mathrm{i}\eta(\boldsymbol{k} \cdot \boldsymbol{\Gamma})f\end{aligned}. \tag{52}$$

Let $\boldsymbol{e}_t = (n_1, n_2, n_3)$ with $n_1^2 + n_2^2 + n_3^2 = 1$, $A_0(\rho) = S_\pm F_0(\rho)$, where

1) As $\boldsymbol{s} = \boldsymbol{\sigma}$, $f(t,\boldsymbol{\rho}) = \zeta, \chi$, one has

$$S_+ = \frac{1}{\sqrt{2(1+n_3)}}\begin{pmatrix} 1+n_3 \\ n_1 + \mathrm{i}n_2 \end{pmatrix}, \quad S_- = \frac{1}{\sqrt{2(1+n_3)}}\begin{pmatrix} -n_1 + \mathrm{i}n_2 \\ 1+n_3 \end{pmatrix}, \quad (\boldsymbol{\sigma} \cdot \boldsymbol{e}_t)S_\pm = \pm S_\pm; \tag{53}$$

2) As $\boldsymbol{s} = \boldsymbol{\tau}$, $f(t,\boldsymbol{\rho}) = \boldsymbol{E'}, \boldsymbol{H'}$, one has

$$S_{+} = S_{-}^{*} = \frac{1}{\sqrt{2}} \begin{pmatrix} n_1 n_3 - i n_2 \\ n_1 - i n_2 \\ n_2 n_3 + i n_1 \\ n_1 - i n_2 \\ -(n_1 + i n_2) \end{pmatrix}, \quad (\boldsymbol{\tau} \cdot \boldsymbol{e}_t) S_{\pm} = \pm S_{\pm}. \tag{54}$$

To lay stress on the spin-orbit coupling interaction, let $\boldsymbol{k} \cdot \boldsymbol{e}_\rho = 0$, i.e., $\cos\Theta = 0$, $\sin\Theta = 1$, using Eqs. (52)-(54), (50) and $\boldsymbol{\Gamma} = \boldsymbol{e}_\rho [\partial A(\rho,t)/\partial \rho]$, one can obtain

$$\eta^2 \omega^2 = \eta^2 \boldsymbol{k}^2 + 2\boldsymbol{\Lambda} \cdot \boldsymbol{\Gamma} + \boldsymbol{\Gamma}^2 + \nabla \cdot \boldsymbol{\Gamma} \pm 2\eta |\boldsymbol{k}| \Lambda, \tag{55}$$

The last term on the right-hand side of Eq. (55) is the gravitational spin-orbit coupling, which implies that, for a given momentum vector of $\boldsymbol{k}$, the spin-orbit couplings of massless Dirac particles and photons in Schwarzschild spacetime are the same (in general, it is $\pm 2\eta |\boldsymbol{k}| \Lambda \sin\Theta$). Using $\boldsymbol{\Gamma} = \boldsymbol{e}_\rho [\partial A(\rho,t)/\partial \rho]$ and $\boldsymbol{e}_\rho = \boldsymbol{\rho}/\rho$, one has

$$\nabla \cdot \boldsymbol{\Gamma} = \partial^2 A(\rho,t)/\partial \rho^2 + 2\partial A(\rho,t)/\rho \partial \rho. \tag{56}$$

Substituting Eqs. (50), (56) and $\boldsymbol{\Gamma} = \boldsymbol{e}_\rho [\partial A(\rho,t)/\partial \rho]$ into Eq. (55), one has

$$\begin{aligned}\eta^2 \omega^2 = \eta^2 \boldsymbol{k}^2 \pm 2\eta |\boldsymbol{k}| \Lambda + (\partial \ln A/\partial \rho)^2 + 2\partial \ln A/\rho \partial \rho \\ + \partial^2 \ln A/\partial \rho^2 - 2\Lambda(\partial \ln A/\partial \rho)\end{aligned}. \tag{57}$$

Let us discuss Eq. (55) or Eq. (57) as follows:

1). As $\rho \to +\infty$ (i.e., $r \to +\infty$, see Eq. (2)), $\eta = (1 - r_s/4\rho)^{-1}(1 + r_s/4\rho)^3 \to 1$, $\Lambda \to 0$, $A(\rho)$ becomes a constant, such that $\boldsymbol{\Gamma} \to 0$, it follows from Eq. (55) that $\omega^2 = \boldsymbol{k}^2$.

2). As $\rho \to r_s/4$ (i.e., $r \to r_s$, see Eq. (2)), $(1 - r_s/4\rho)^{-1} = \eta(1 + r_s/4\rho)^{-3} = \eta/8$, and $\Lambda = \eta/4r_s$, where $\eta \to +\infty$, it follows from Eq. (57) that

$$\omega^2 = \boldsymbol{k}^2 \pm |\boldsymbol{k}|/2r_s + \lim_{\rho \to r_s/4} F(t,\rho)/\eta^2, \tag{58}$$

where

$$F(t,\rho) = (\partial \ln A/\partial \rho)^2 + 2\partial \ln A/\rho \partial \rho + \partial^2 \ln A/\partial \rho^2 - \eta(\partial \ln A/\partial \rho)/2r_s. \tag{59}$$

By imposing *the slowly-varying envelope approximation* on Eq. (51), one can take the approximation of $\lim_{\rho \to r_s/4} F(t,\rho)/\eta^2 \approx 0$. Then we have, for $\rho \to r_s/4$,

$$\omega^2 = |\bm{k}|^2 \pm |\bm{k}|/2r_s,  \qquad (60)$$

The second term on the right-hand side of Eq. (60) comes from the contribution of the spin-orbit coupling. More generally, it is $\pm|\bm{k}|\sin\Theta/2r_s$. When a Schwarzschild black hole shoots off a Dirac particle or a photon tangentially, if the probability of the particle escaping from the black hole does not vanish, one has $\omega^2 > 0$, for the moment Eq. (60) implies that $|\bm{k}| > 1/2r_s$. On the other hand, for $\rho \to r_s/4$ Eq. (60) implies that

$$\omega = \pm\omega_\pm, \quad \omega_\pm = \sqrt{|\bm{k}|^2 \pm |\bm{k}|/2r_s},  \qquad (61)$$

where $\omega_+$ and $\omega_-$ respectively represent the energies of massless Dirac particles with the spin projections of $\pm 1/2$, or respectively represent the energies of photons with the spin projections of $\pm 1$. That is, because of the gravitational spin-orbit coupling, there is a splitting of energy levels. As we know, the helicity of a particle describes the spin orientation with respect to the direction of the particle's motion. However, in the spin-orbit coupling, the spin projection of a particle describes the spin orientation with respect to the direction of an orbital angular momentum (in our case, along the direction of $(\bm{e}_\rho \times \bm{k})$).

As we know, information about the collapsed matter in a black hole will be lost if Hawking radiations are truly thermal, which is inconsistent with the unitarity of quantum mechanics, and then presents a serious obstacle for developing theories of quantum gravity. This is the so-called "information loss paradox". Many investigations on the question of whether information is lost in black holes have been presented [19-25], but none is capable of successfully ending the dispute.

In view of the fact that free (i.e., non-self-interacting) quantum fields in curved spacetime are equivalent to the quantum fields of Minkowski spacetime interacting with a gravitational field, the Hawking radiations of Dirac particles or photons in the presence of the gravitational spin-orbit couplings are the same as the ones of free Dirac particles or photons in Schwarzschild spacetime. Therefore, because of the spin-orbit couplings, when Dirac particles or photons coming from the Hawking radiation near the event horizon of a Schwarzschild black hole are emitted along the tangential direction of the Schwarzschild black hole, their energies are given by Eq. (61). Let $T = T_H$ denote the Hawking temperature, $N_d$ and $N_p$ denote the Hawking thermal spectrums for Dirac particles and photons, respectively, one has

$$N_d(\omega) = [\exp(\omega/k_B T_H) + 1]^{-1}, \quad N_p(\omega) = [\exp(\omega/k_B T_H) - 1]^{-1}, \qquad (62)$$

where $k_B$ is the Boltzmann constant. Because of the gravitational spin-orbit couplings, for each given momentum vector of $\boldsymbol{k}$, one has $\omega = \omega_\pm$, i.e., there is an energy-level splitting between two states with different spin projections and the same momentum vector. It follows from Eq. (62) and $\omega_+ > \omega_-$ that

$$N_d(\omega_+) < N_d(\omega_-), \quad N_p(\omega_+) < N_p(\omega_-). \qquad (63)$$

Therefore, for a given momentum vector of $\boldsymbol{k}$, Dirac particles or photons coming from the Hawking radiations are partially polarized (provided that their original momentums are not parallel to the radial direction, i.e., $\sin\Theta \neq 0$), rather than completely disorder, which might encode the information and escape the black hole.

This work was supported by the Fundamental Research Funds for the Central Universities (Grant No.ZYGX2010X013).

* zywang@uestc.edu.cn